\documentclass[twocolumn,showpacs,preprintnumbers,amsmath,amssymb]{revtex4}
\usepackage{graphicx}
\begin{document}


\title{Competition between the BCS superconductivity and ferromagnetic spin fluctuations in MgCNi$_3$}

\author{L. Shan, Z.Y. Liu }
\author{Z.A. Ren, G.C. Che}
\author{H.H. Wen}\email{hhwen@aphy.iphy.ac.cn}

\affiliation{ $^1$National Laboratory for Superconductivity, Institute of Physics,
Chinese Academy of Sciences, P.O. Box 603, Beijing 100080, China}


\date{\today}

\begin{abstract}
The low temperature specific heat of the superconductor MgCNi$_3$ and a
non-superconductor MgC$_{0.85}$Ni$_3$ is investigated in detail. An additional
contribution is observed from the data of MgCNi$_3$ but absent in MgC$_{0.85}$Ni$_3$,
which is demonstrated to be insensitive to the applied magnetic field even up to 12
Tesla. A detailed discussion on its origin is then presented. By subtracting this
additional contribution, the zero field specific heat of MgCNi$_3$ can be well described
by the BCS theory with the gap ratio ($\Delta/k_BT_c$ ) determined by the previous
tunneling measurements. The conventional $s$-wave pairing state is further proved by the
magnetic field dependence of the specific heat at low temperatures and the behavior of
the upper critical field.

\end{abstract}


\maketitle

Since the discovery of the new intermetalic perovskite superconductor MgCNi$_{3}$
\cite{HeT2001}, plenty of efforts have been focused on the superconducting pairing
symmetry in this material because its conduction electrons are derived predominantly from
Ni which is itself a ferromagnet \cite{RosnerH2002,DugdaleSB2001,SinghDJ2001,ShimJH2001}.
However, up to now, there is still not a consensus on this issue. The measured
penetration depth \cite{ProzorovR2003}, critical current behavior \cite{YongDP2003} and
earlier tunneling spectra \cite{MaoZQ2003} suggested an unconventional superconductivity,
the later tunneling data \cite{ShanL2003t} supported the $s$-wave pairing symmetry and
gave a reasonable interpretation on the contradiction to the result in
Ref\cite{MaoZQ2003}. The $s$-wave pairing has also been demonstrated by the $^{13}$C NMR
experiments \cite{SingerPM2001} and the specific heat measurements
\cite{HeT2001,LiSY2001,WalteA2002,WalteA2004,MaoZQ2003,ShanL2003h}. To our knowledge, all
the previous reports on the specific heat of MgCNi$_{3}$
\cite{HeT2001,LiSY2001,WalteA2002,WalteA2004,MaoZQ2003,ShanL2003h} were characterized in
the framework of a conventional phonon-mediated pairing. However, there is an obvious
deviation of the experimental data from the prediction of BCS theory in the low
temperature \cite{LinJY2003,MaoZQ2003}, i.e., the entropy conservation rule is not
satisfied. Such deviation has been interpreted by the presence of unreacted Ni impurities
in Refs\cite{LinJY2003,MaoZQ2003}, whereas it is still prominent in the samples without
Ni impurities \cite{WalteA2004}. On the other hand, strong spin fluctuations have been
observed in MgCNi$_{3}$ by NMR experiment \cite{SingerPM2001}, which is suggested to be
able to severely affect the superconductivity in MgCNi$_{3}$
\cite{RosnerH2002,SinghDJ2001,SingerPM2001,ShanL2003h,HaywardMA2001} or even induce some
exotic paring mechanism \cite{RosnerH2002}. Consequently, the behavior of the specific
heat will inevitably be changed by the spin fluctuations. Therefore, before a real
pairing mechanism being concluded from the specific heat data, we have to carefully
investigate how the ferromagnetic spin fluctuations contribute to the specific heat of
MgCNi$_{3}$.

In this work, we elaborate on the specific heat ($C$) of MgC$_x$Ni$_{3}$ system both in
normal state and superconducting state. A low temperature upturn is clearly distinguished
in the $C/T$ vs $T^2$ curves and found to be insensitive to the applied magnetic field.
By doing some quantitative analysis, we present the evidence of most possible mechanisms
responsible for this upturn. After subtracting this additional contribution, a well
defined BCS-type electronic specific heat is extracted. The temperature dependence of the
upper critical field and the field dependence of the low temperature specific heat also
supports such conventional BCS superconductivity in MgCNi$_{3}$. These analyses indicate
that although the spin fluctuations may suppress the pairing strength in MgCNi$_{3}$, the
superconductivity is certainly not induced by any exotic mechanism.

Poly-crystalline samples of MgC$_x$Ni$_3$ were prepared by powder metallurgy method.
Details of the preparation were published previously \cite{RenZA2002}. The superconductor
MgCNi$_{3}$ has a $T_c$ of $6.7K$ and the non-superconductor MgC$_{0.85}$Ni$_3$ was
synthesized by continually reducing the carbon component until the diamagnetism was
completely suppressed. The heat capacity data presented here were taken with the
relaxation method \cite{Bachmann1972} based on an Oxford cryogenic system Maglab in which
the magnetic field can be achieved up to 12 Tesla. Details of the sample information and
the measurements can be found in recent report \cite{ShanL2003h}. It should be emphasized
here that the Cernox thermometer used for calorimetry has been calibrated at 0, 1, 2, 4,
8 and 12 Tesla, and the calibration for the intermediate fields is performed by an
interpolation using the result of the adjacent fields. Therefore, any prominent field
dependence of the specific data should reflect the intrinsic properties of the measured
sample.

\begin{figure}[t]
\includegraphics[scale=0.8]{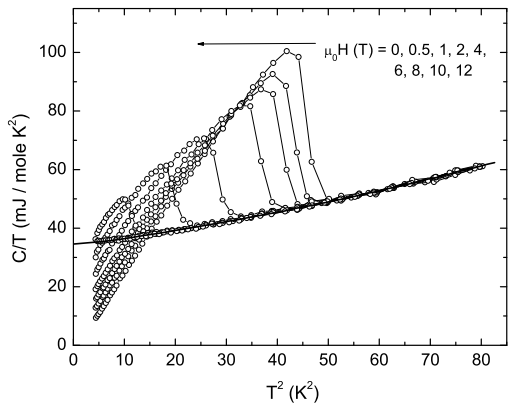}
\caption{\label{fig:fig1} The low-temperature specific heat of MgCNi$_{3}$ at various
magnetic fields from 0T to 12T. The thick solid line denotes the universal background
including all the normal state data for different fields.}
\end{figure}

In general, the low temperature specific heat $C(T,H)$ of a superconductor consists four
main contributions by neglecting the component of the nuclear moments
\cite{Specificheat,Emerson1999}, each has a different dependence on $T$ and two of which
depend on $H$, also in different ways,
\begin{equation}
C(H,T)=C_{mag}(H,T)+C_{DOS}(H,T)+\gamma_0T+C_{ph}(T)
\label{eq:one}
\end{equation}
where $\gamma_0T$ represents a spare zero-field linear term and
$C_{ph}$ is due to the lattice or phonon contribution. The Debye
phonon specific heat $C_{ph}=\beta T^3$ can usually describe the
lattice contribution at low temperatures. However, the departure
from $T^3$-law has often been observed, which is due to the fact
that the density of modes of the phonon in real solid does not
follow the assumed $\omega^2$-law, here $\omega$ is the angular
frequency of a harmonic wave associated with the lattice
vibration. In such case, the deviations may be expanded in higher
order terms such as $T^5$, $T^7$, etc.. The $H$-dependent terms in
Eq.~(\ref{eq:one}), i.e., $C_{mag}(H)$ and $C_{DOS}(H)$, are the
contributions associated with magnetism and the electronic density
of states ( DOS ), respectively. If there is no magnetism
associated contribution, the normal state specific heat at low
temperature can be approximatively described as
$C_n(T)=\gamma_nT+\beta T^3$ in the framework of metal theory.
Therefore, a linear relation can be obtained by plotting the
normal state data as $C_n(T)/T$ vs $T^2$, and its intercept and
slope correspond to $\gamma_n$ and $\beta$, respectively.

\begin{figure}[t]
\includegraphics[scale=0.8]{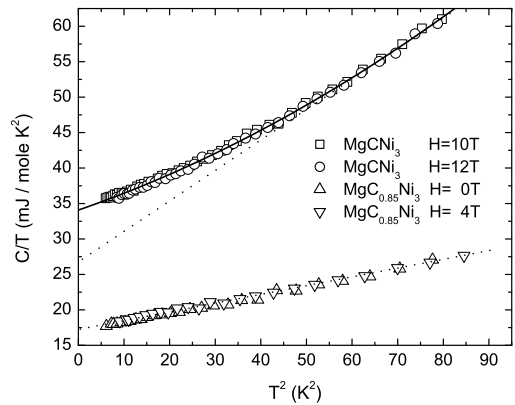}
\caption{\label{fig:fig2}A plot of $C/T$ vs $T^2$ for MgCNi$_{3}$ and
MgC$_{0.85}$Ni$_{3}$ in different magnetic fields. The data of MgC$_{0.85}$Ni$_{3}$ can
be well defined by a straight line while that of MgCNi$_{3}$ remarkably deviate the
linearity. The dashed line is a linear extrapolation of the high temperature data of
MgCNi$_{3}$ and the upper solid line is the theoretical fit considering higher order
phonon contributions.}
\end{figure}
The low temperature specific heat at various magnetic fields up to 12 Tesla is plotted as
$C(T)/T$ vs $T^2$ in Fig.~\ref{fig:fig1}. Two important features should be emphasized
here. First, all the normal state data at various magnetic fields merge into one
\cite{ShanL2003h}, which is consistent with the results reported by other groups
\cite{WalteA2004,LinJY2003}. Second, this common normal state background remarkably
deviates from the linear relation as discussed above. In Fig.~\ref{fig:fig2}, only the
normal state data are re-plotted in an magnified scale. In order to survey the normal
behavior at very low temperature, the magnetic field up to 12 Tesla was applied in
measurements, which exceeds the highest upper critical field of our sample and is 4 Tesla
higher than that used by other groups \cite{WalteA2004,LinJY2003}. The specific heat of
the non-superconductor MgC$_{0.85}$Ni$_3$ is also presented in Fig.~\ref{fig:fig2} as a
comparison. It is obvious that the perfect linear relation of $C(T)/T$ vs $T^2$ is
satisfied for MgC$_{0.85}$Ni$_3$, which is a striking contrast with the case of
MgCNi$_{3}$.

\begin{figure}[t]
\includegraphics[scale=0.8]{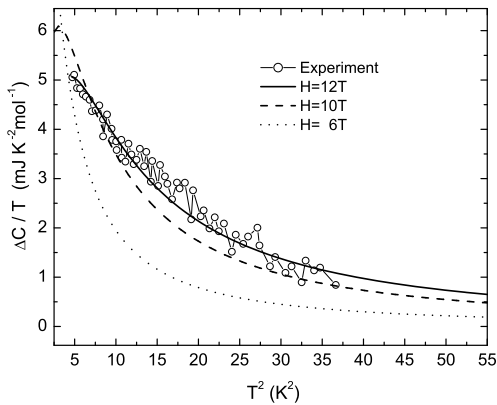}
\caption{\label{fig:fig3} Comparison between the field-insensitive low-temperature upturn
in the specific heat of MgCNi$_{3}$ ( $\Delta C=C-\gamma T-\beta T^3$) and the calculated
field dependent Schottky anomaly.}
\end{figure}

The obvious upturn in the low temperature $C(T)/T$ vs $T^2$ curves of MgCNi$_{3}$ can not
be associated with Ni impurites since the X-ray diffraction pattern shows no indication
for Ni impurites \cite{ShanL2003h}. To say the least, if there is still extreme small
content of Ni impurites leading to the prominent low temperature upturn of $C/T$, the
field dependence of its specific heat should also be obvious, which is clearly
inconsistent with our experimental results. Moreover, if this upturn is due to the excess
free Ni in MgCNi$_{3}$, it should also be observed in MgC$_{0.85}$Ni$_{3}$ because of the
similar process of synthesizing these two samples. Quantitatively, taking the data from
references \cite{DixonM1965,CaudronR1973} yields for 10\% of superfluous Ni an upturn
which is at least two orders of magnitude smaller than the observed one. Therefore, the
contribution of the excessive Ni can be neglected comparing with the whole specific heat.
Furthermore, the possible Schottky anomaly is presented in Fig.\ref{fig:fig3}, its field
dependence is obviously too strong to compare with our experimental result ( nearly field
independent ).

It is found that this upturn can be well fitted if the above mentioned $T^5$ term is
considered ( see the upper solid line in Fig.~\ref{fig:fig2} ). In other words, the
departure from the $T^3$ behavior may be due to the non-Debye phonon DOS, which is
consistent with the notable difference of the Debye temperature between MgCNi$_{3}$ and
MgC$_{0.85}$Ni$_3$ \cite{ShanL2003h}. If the electron-phonon coupling is indeed the
origin of superconductivity in MgCNi$_{3}$, it is reasonable to associate the
disappearance of superconductivity in MgC$_{0.85}$Ni$_3$ with the remarkable difference
of its phonon DOS from that of MgCNi$_{3}$. However, some careful work is needed to
understand such obvious difference of phonon structure between these two samples, since
they have similar crystal lattices and chemical components.

\begin{figure}[t]
\includegraphics[scale=0.8]{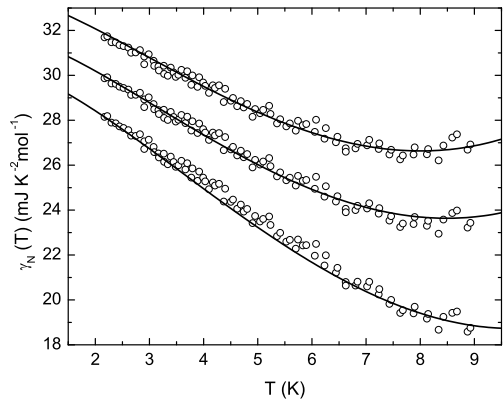}
\caption{\label{fig:fig4}Electronic specific heat $\gamma_n(T)$ versus $T$ of MgCNi$_{3}$
in the normal state ( three different $\beta$-values are selected in order to avoid
artificial errors. ). The solid lines are theoretical fits to spin-fluctuation model. All
the curves except the top two are shifted downwards for clarity. }
\end{figure}

Another possible explanation of the above mentioned low temperature upturn is the
existence of strong spin fluctuations due to the higher DOS at fermi energy ( $N(E_F)$ )
of MgCNi$_{3}$ than that of MgC$_{0.85}$Ni$_{3}$ \cite{ShanL2003h}, consequently, the
coupling between the electrons and spin fluctuations in MgCNi$_{3}$ should also be
stronger. The ferromagnetic spin fluctuations have been demonstrated by NMR experiments
\cite{SingerPM2001}. Doniach and Engelshberg \cite{DoniachS1966} and Berk and Schrieffer
\cite{BerkNF1966} showed that the absorption and re-emission of spin fluctruations
renormalizes the electronic self-energy, leading to an enhanced effective mass at low
temperatures. This effect manifests itself as a low-temperature enhancement of the
electronic specific-heat coefficient, $\lambda_{sf}$, which depends on temperature as
$T^2ln(T/T_{sf})$ (here $T_{sf}$ is the characteristic spin-fluctuation temperature) at
low temperature. Considering the presence of ferromagnetic spin fluctuations, the normal
state specific heat of MgC$_{x}$Ni$_3$ can be expressed as follows,
\begin{equation}
C_n(H=0,T)=A[1+\lambda_{ph}+\lambda_{sf}(T)]T+\gamma_0T+\beta
T^3\label{eq:two}
\end{equation}
where $\beta T^3$ are the contributions of phonon excitations,
$\lambda_{sf}T$ and $\lambda_{ph}T$ represent the contributions of
effective mass renormalization due to the electron-spin
fluctuation coupling and the electron-phonon coupling,
respectively, and $A$ is a constant correlated with $N(E_F)$. It
can be seen from Eq.~(\ref{eq:two}) that the deviation from the
linear dependence of $C(T)/T$ on $T^2$ is due to the temperature
dependence of $\lambda_{sf}$. Moreover, B$\acute{e}$al-Monod, Ma,
and Fredkin \cite{Beal-MonodMT1968} have estimated the shift
$\delta C/T$ caused by an applied field $H$ to be
\begin{equation}
\delta C/T\approx 0.1(\frac{\mu H}{k_{B}T_{sf}})^2\frac{S}{lnS}
\label{eq:twoB}
\end{equation}
where $S$ is Stoner factor. Eqs.~(\ref{eq:two}) and
~(\ref{eq:twoB}) indicate that the possible magnetic field
dependence of the normal state specific heat is completely
determined by the spin fluctuations. For simplicity,
Eq.~(\ref{eq:two}) can be rewritten as
$C_n(H=0,T)=\gamma_n(T)T+\beta T^3$, in which
$\gamma_n(T)=A[1+\lambda_{ph}+\lambda_{sf}(T)+\gamma_0/A]$.
Therefore, the $\gamma_n\sim T$ relation directly reflects the
temperature dependence of $\lambda_{sf}$. In Fig.~\ref{fig:fig4},
we present the determined $\gamma_n(T)$ by selecting various
$\beta$-values. Fitting the $\gamma_n(T)$ relations to the formula
of $A(1+BT^2ln(T/T_{sf}))$ yields $T_{sf}$ varying from 13 to 16K.
By inserting the determined $T_{sf}$, calculated Stoner factor $S$
\cite{RosnerH2002} and the highest field value in our measurements
into Eq.~(\ref{eq:twoB}), we can estimate the shift $\delta C/T$
caused by the applied field to be less than 2\%, which is in
agreement with our experimental results. However, if this
explanation is correct, we must understand the collapse of the
entropy conservation around $T_c$ caused by considering such
additional electronic specific heat, as discussed below.
Therefore, the specific-heat contribution of the spin fluctuations
themselves may be another candidate responsible for the low
temperature upturn in specific heat of MgCNi$_3$.

\begin{figure}[top]
\includegraphics[scale=0.95]{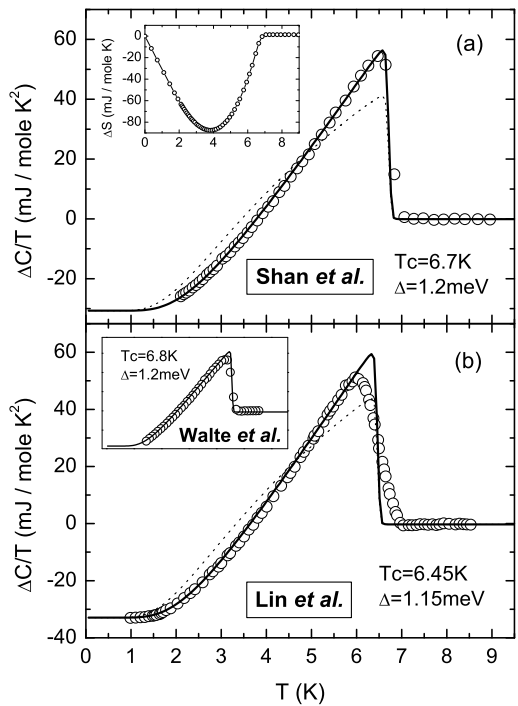}
\caption{\label{fig:fig5} Fitting the specific heat data
($C_{H=0}-C_n$) measured by different groups
\cite{WalteA2004,LinJY2003,ShanL2003h} to BCS model. The
experimental data are denoted by open circles and the fits to
$\alpha$-model are denoted by solid lines. The dotted lines are
fits to the original BCS model. The inset in (a) is the entropy
difference by integration of $\Delta C/T$ as presented in (a). }
\end{figure}

Despite the true mechanism of the low-temperature upturn of $C/T$, this additional
specific heat contribution should be regarded as a part of the normal-state background of
the superconducting specific heat below the upper critical field $H_{c2}(T)$. In earlier
analysis to the specific heat data \cite{LiSY2001,MaoZQ2003,ShanL2003h}, this additional
part of background has been neglected more or less below $H_{c2}(T)$. We point out here
that neglecting this additional contribution will lead to the collapse of the entropy
conservation as reported in references \cite{MaoZQ2003,LinJY2003}. This opinion is
motivated by the subsequent analysis. As shown in Fig.~\ref{fig:fig5}(a), the normal
state background ( as shown in Fig.~\ref{fig:fig2} ) has been subtracted from the
zero-field specific heat data, the entropy difference $\Delta S(T)=\int_0^{T}dT(\Delta
C/T)$ is presented in the inset of Fig.~\ref{fig:fig5}(a), here $\Delta C=C_{H=0}-C_n$.
It is found that the entropy conservation is then well satisfied, indicating that the
remainder is the contribution of superconducting state. Such analysis has also been
applied on the data measured by W\"{a}lte \textit{et al.} \cite{WalteA2004} and Lin
\textit{et al.} \cite{LinJY2003}, respectively, the entropy is also conserved and the low
temperature anomaly as mentioned in Ref.\cite{LinJY2003} completely vanishes.

When the superconductivity in MgCNi$_{3}$ is investigated, the spin fluctuations can not
be neglected because it may compete with superconductivity \cite{ShanL2003h,SinghDJ2001}
or even lead to an exotic pairing mechanism other than the conventional s-wave
\cite{RosnerH2002}. If the phonon-mediated pairing wins in competing with the spin
fluctuations and hence the effect of the spin fluctuations only suppress the
electron-phonon coupling ( or pairing-strength ) \cite{ShanL2003h}, the so-called
$\alpha$-model \cite{PadamseeH1973} based on BCS theory should be a good choice to
describe the measured thermodynamic parameters. Comparing with the original BCS-model,
the only adjustable parameter in this $\alpha$-model is the gap ratio $\Delta(0)/k_BT_c$.
This model has been successfully applied to strong coupling systems such as Pb and Hg.
Then we try to fit the superconducting part of the zero-field specific heat to the
$\alpha$-model, the results are presented in Fig.~\ref{fig:fig5}. All the data can be
well described by this revised BCS model with the best fitting parameters ( i.e.,
$\Delta$ and $T_c$ ) listed in Table.~\ref{tab:table1}.
\begin{table}
\caption{\label{tab:table1}Fits to BCS model for zero-field
specific heat.}
\begin{ruledtabular}
\begin{tabular}{lccc}
Groups & $\Delta$ (meV)  & $T_c$ (K)  & $2\Delta/k_BT_c$ \\
\hline
Shan \textit{et al.} \cite{ShanL2003h}       & 1.20            &6.70        &4.15  \\
W\"{a}lte \textit{et al.} \cite{WalteA2004}  & 1.20            &6.80        &4.10  \\
Lin \textit{et al.} \cite{LinJY2003}         & 1.15            &6.45        &4.14  \\

\end{tabular}
\end{ruledtabular}
\end{table}
These fits yield a gap ratio $\Delta(0)/k_BT_c\approx 2.06$, corresponding to the maximum
gap $\Delta(0)\approx 1.2$meV which is in good agreement with our previous tuneling
measurements \cite{ShanL2003t}. From the above discussions, we can conclude that the
coexistence and competition of spin fluctuations and phonons does not change the
phonon-mediated pairing mechanism of MgCNi$_3$.

In order to further verify this picture, we investigate the field dependence of the low
temperature specific heat of MgCNi$_3$. It is known that the electronic specific heat in
magnetic fields can be expressed by $C_{el}(T,H)=C_{el}(T,H=0)+\gamma(H)T$. The magnetic
field dependence of $\gamma(H)$ is associated with the form of the gap function of the
superconductor. For example, in a superconductor with line nodes in the gap function, the
quasiparticle DOS ( $N(E)$ ) rises linearly with energy at the Fermi level in zero field,
$N(E)\propto|E-E_F|$, which results in a contribution to the specific heat
$C_{DOS}=\alpha T^2$ \cite{Kopnin1996}. In the mixed state with the field higher than a
certain value, the DOS near the Fermi surface becomes finite, therefore the quadratic
term $C_{DOS}=\alpha T^2$ will disappear and be substituted by the excitations from both
inside the vortex core and the de-localized excitations outside the core. For d-wave
superconductors with line nodes in the gap function, Volovik {\it et al.}
\cite{Volovik1993} pointed out that in the mixed state, supercurrents around a vortex
core cause a Doppler shift of the quasi-particle excitation spectrum. This shift has
important effects upon the low energy excitation around the nodes, where its value is
comparable to the width of the superconducting gap. For $H >> H_{c1}$, it is predicted
that $N(E_F)\propto H^{1/2}$ and $C_{DOS} =\Delta\gamma (H)T = ATH^{1/2}$ at low
temperatures \cite{Volovik1993}. This prediction has been well proved for hole doped
cuprates \cite{Specificheat}. Whereas in a conventional $s$-wave superconductor, the
specific heat in the vortex state is dominated by the contribution from the localized
quasi-particles in the vortex cores. From the Bogoliubov equations assuming
noninteracting vortices, the DOS associated with the bound excitations is derived as
$N(E)\propto B(H)$ \cite{FetterAL1969} hence the contribution of the vortex cores to the
specific heat is $C_{DOS}\propto B(H)T$ \cite{RamirezAP1996,SonierJE1999}. It is also
theoretically derived that the experimentally observed downward curving $C(H)$ is caused
by the flux line interactions near $H_{c1}$ and the possible expansion of the vortex
cores \cite{RamirezAP1996,SonierJE1999}. According to the above discussions, the specific
heat coefficient $\gamma(H)$ of conventional $s$-wave superconductor should linearly
depend on the magnetic field well above $H_{c1}$.

\begin{figure}[]
\includegraphics[scale=0.8]{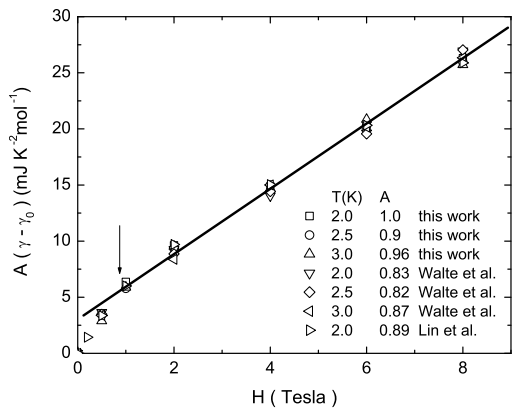}
\caption{\label{fig:fig6}The magnetic field dependence of the specific heat coefficient
$\gamma(H)-\gamma(0)$ at low temperatures.}
\end{figure}

Fig.~\ref{fig:fig6} shows the field dependence of $\gamma(H)-\gamma(0)$ of MgCNi$_3$
below 3K. The data reported by different groups \cite{ShanL2003h,WalteA2004,LinJY2003}
merge into each other by timing a prefactor $A$ close to unity. It is found that $\gamma$
linearly depends on $H$ above 0.5T and persists up to 8T which is close to the upper
critical field of MgCNi$_3$. The legible linearity of $\Delta\gamma\sim H$ relation at
higher field and its negative curvature below 0.5T are in good agreement with the above
mentioned behaviors of conventional $s$-wave superconductors. It may be argued that the
low temperature limit of about 2K in our measurements is not low enough to distinguish
the $d$-wave's $\Delta\gamma\sim H^{1/2}$-law. However, it should be emphasized that the
observed $\Delta\gamma\sim H$ relation is nearly universal at low temperatures below
upper critical field, which is very similar to the behavior of $V_3Si$
\cite{RamirezAP1996}, a typical conventional $s$-wave superconductor.

\begin{figure}[t]
\includegraphics[scale=0.8]{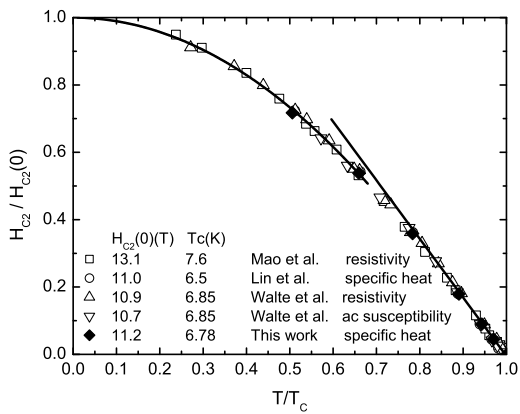}
\caption{\label{fig:fig7}The comparison of the temperature dependence of the upper
critical field $H_{c2}(T)$ with the BCS-like descriptions in lower and higher temperature
limits.}
\end{figure}

Finally, we compare the temperature dependence of the upper critical field $H_{c2}(T)$
with the prediction of BCS theory in which the $H_{c2}(T)$ can be expressed as follows,
\begin{subequations}
\label{eq:whole}
\begin{equation}
H_{c2}(T)\approx1.74H_c(0)(1-T/T_c)\;\;\;\;\; (T_c-T\ll
T_c)\label{subeq:1}
\end{equation}
\begin{equation}
H_{c2}(T)\approx H_{c2}(0)[1-1.06(T/T_c)^2]\;\;\;\;\;\; (At\;low\;
T)\label{subeq:2}
\end{equation}
\end{subequations}
As shown in Fig.~\ref{fig:fig7}, the best fitting to BCS model is
denoted by solid lines. At lower temperature, the experimental
data can be well described by Eq.(\ref{eq:whole}b). For the higher
temperature near $T_c$, a prefactor of 1.65 is obtained instead of
the theoretical prediction of 1.74 as expressed in
Eq.(\ref{eq:whole}a). Nonetheless, the BCS model is still a
preferred description for $H_{c2}(T)$ of MgCNi$_3$ considering the
stronger electron-phonon coupling and the presence of
ferromagnetic spin fluctuations.

In summary, we have investigated the specific heat data of MgC$_x$Ni$_3$ system. A
remarkable field independent contribution is found in MgCNi$_3$, reflecting the departure
of normal-state specific heat from $T^3$-law. By removing this contribution, the
zero-field data is well described by the $\alpha$-model ( a slightly revised BCS model ).
The conventional $s$-wave superconductivity is further supported by the linear field
dependence of specific heat coefficient $\gamma(H)$ and the BCS-like temperature
dependence of upper critical field $H_{c2}(T)$. It is then concluded that, although
electron-magnon ( spin fluctuations ) coupling coexists and competes with electron-phonon
coupling effect in MgCNi$_3$, it only acts as pair breakers while does not induce a new
exotic superconductivity.

Note added: Most recently, the carbon isotope effect in superconducting MgCNi$_{3}$
observed by T. Klimczuk and R.J. Cava indicates that carbon-based phonons play a critical
role in the presence of superconductivity in this compound \cite{CavaRJ2004}.

\begin{acknowledgments}
This work is supported by the National Science Foundation of China, the Ministry of
Science and Technology of China, and Chinese Academy of Sciences with the Knowledge
Innovation Project.
\end{acknowledgments}


\end{document}